\documentstyle[12pt]{article}
\input epsf
\newwrite\ffile\global\newcount\figno \global\figno=1

\def\writedef#1{}
\def\figin{\epsfcheck\figin}\def\figins{\epsfcheck\figins}
\def\epsfcheck{\ifx\epsfbox\UnDeFiNeD
\message{(NO epsf.tex, FIGURES WILL BE IGNORED)}
\gdef\figin##1{\vskip2in}\gdef\figins##1{\hskip.5in}
\else\message{(FIGURES WILL BE INCLUDED)}%
\gdef\figin##1{##1}\gdef\figins##1{##1}\fi}
\def\figinsert{}
\def\ifig#1#2#3{\xdef#1{fig.~\the\figno}
\writedef{#1\leftbracket fig.\noexpand~\the\figno}%
\figinsert\figin{\centerline{#3}}\medskip\centerline{\vbox{\baselineskip12pt
\advance\hsize by -1truein\center\footnotesize{  Fig.~\the\figno.} #2}}
\bigskip\endinsert\global\advance\figno by1}
\def\endinsert{}

\begin{document}

\begin{flushright}
RU--98--40 \\
\today 
\end{flushright}

\begin{center}
\bigskip\bigskip
{\bf  The Spectrum and Effective Action of SUSY Gluodynamics}
\footnote{Talk given  by G. Gabadadze at the International
Seminar ``Quarks-98'', May 18-24, 1998, Suzdal, Russia.}

\vspace{0.3in}      

{G.~R. ~Farrar\footnote{New address: Physics Department, New York 
University, USA.}
,~ G. Gabadadze,  ~M. Schwetz}
\vspace{0.2in}

{\baselineskip=14pt
Department of Physics and Astronomy, Rutgers University \\
Piscataway, New Jersey 08855, USA}

\vspace{0.2in}
\end{center}

\vspace{0.2in}

{\small   We study the low-energy spectrum of SUSY gluodynamics  
using the generating functional for Green's functions
of composite fields. Two dual formulations of the generating 
functional approach  
are  given. Masses of the bound states are calculated and 
mixing patterns are discussed. Mass splittings of pure gluonic states,
in the case when supersymmetry is softly broken,  
are consistent with predictions of conventional Yang-Mills theory. 
The  results can be tested in lattice simulations of the 
SUSY Yang-Mills model.} 
\vspace{1.5cm}

Supersymmetric gluodynamics, the theory of gluons and gluinos,
is an extremely useful testing ground for various
nonperturbative phenomena occuring in conventional QCD.  
The Witten index of the $SU(N_c)$ SUSY gluodynamics
equals to $N_c$ \cite {WittenN}. Thus, the ground
state of the model consists of at least $N_c$ different 
vacua parametrized by the imaginary phase of a nonzero gluino condensate 
\cite {WittenN, Condensate}. The different 
vacua are related by discrete $Z_{2N_c}$ transformations
of gluino fields. Once one of the $N_c$ vacua is 
chosen, the $Z_{2N_c}$ symmetry group spontaneously breaks down   
to the $Z_2$ subgroup. 

In analogy with QCD, one expects that in each of those vacua, 
the spectrum of the   model consists 
of colorless bound states of gluinos and gluons.
Among those are: pure gluonic bound states (glueballs), 
gluino-gluino mesons  and gluon-gluino composites. 
These states fall into the lowest-spin representations 
of the $N=1$ SUSY algebra written in the basis of parity eigenstates
\cite {FGS1, FGS2}. The masses and mixings  of these bound states can  be 
given within the effective Lagrangian approach.  
The effective action for $N=1$ SYM was 
proposed by Veneziano and Yankielowicz (VY) \cite {VY}. 
The VY action \cite {VY} involves fields for gluino-gluino 
and gluino-gluon bound states.  However, it does not include dynamical
degrees of freedom which would  correspond to pure gluonic 
composites (glueballs). 

At this stage  we would like to make a digression and 
comment on the  physical meaning of the VY 
effective action. This is not an  effective action 
in the Wilsonian sense. 
In ref. \cite {Shore} the VY action 
was constructed  as a generating functional of one-particle-irreducible (1PI)
Green's functions \cite {GoldstoneSalamWeinberg}. 
That means that  the VY action, being written in
terms of 
composite colorless fields  of  SYM theory,  can  be used 
to calculate  various Green's functions of those composite variables. 
Performing those calculations, however, one is not supposed to 
take into account
diagrams  with  composite fields  propagating in  virtual loops.
Loop effects  are  already included 
in effective  vertices and
propagators  occuring in the action. 
The simplest kind   of  Green's function  one might be 
interested in is a two point correlator. As we mentioned above, 
the composite operators
entering the VY action are the interpolating 
fields for the bound states of $N=1$ SYM theory. 
Thus, a two point correlator (or simply a 
propagator)  of those fields can be used to determine  the
mass of the corresponding bound state. 
Hence, the effective action,  
or more exactly the generating functional of 1PI diagrams, 
which we deal with in this paper 
can readily be used to deduce masses of composite bound states of the 
theory. In what follows the effective action 
(and effective Lagrangian)
discussed will be understood in the sense specified  above. 

In this report we describe briefly how one can generalize 
the VY Lagrangian in order  to  incorporate 
the glueball states into the description \cite {FGS1, FGS2}. 

The classical action of  $N=1$ SYM theory is invariant  
under chiral, scale 
and superconformal transformations. 
Once quantum effects are taken into account, 
these symmetries are 
broken by the  chiral, scale and superconformal  anomalies respectively.
Composite operators that appear  in the expressions for the 
anomalies can be gathered into   
a composite chiral supermultiplet ${\rm Tr} W^{\alpha}W_{\alpha}$ 
\cite {WessZumino} 

The effective action of the model can be a functional of the
superfield $S$ 
\begin{eqnarray}
S\equiv { \beta (g) \over 2 g} \Big \langle 
{\rm Tr} W^{\alpha}W_{\alpha}\Big \rangle _Q\equiv
A(y)+\sqrt{2} \theta \Psi(y) + \theta^2 F(y),  \nonumber
\end{eqnarray}
where the VEV is defined for nonzero value of an external 
(super)source $Q$ \cite {Shore}.
$\beta(g)$ stands for   the SYM beta function which is known 
exactly \cite {beta}. The lowest component
of the $S$ superfield, $A$, is bilinear in gluino fields and has the quantum
numbers of the scalar and pseudoscalar gluino-gluino  bound states.  The
fermionic component in $S$ is related to  the gluino-gluon composite and
the $F$ component of the chiral superfield includes operators
corresponding to both the scalar and 
pseudoscalar glueballs ($G_{\mu\nu}^2$ and 
$G_{\mu\nu}{\tilde G^{\mu\nu}}$ respectively). 

Assuming that the effective action 
of the 
model can be written in terms of the single  superfield $S$,   and 
requiring also that the effective action respects all the global
continuous symmetries and 
reproduces the anomalies
of the SYM theory,   one derives the Veneziano-Yankielowicz 
effective superpotential \cite {VY} 
\begin{eqnarray}
{\cal W}_{\rm VY}(S)=
\gamma ~S~ {\rm ln} {S\over e \mu^3},   
\label{VYsup}
\end{eqnarray}
where $\gamma \equiv - (N_c g/16 \pi^2 \beta(g))>0$, $\mu$ stands for
the dimensionally transmuted scale of the model and $e\simeq 2.71$. 
If one uses this superpotential along with the simplest 
K\"ahler potential $(S^+S)^{1/3}$, one finds that no glueball 
operators are present in the Lagrangian. Indeed, 
all the glueball fields enter through the $F$ components
of the superfield $S$. These components have no dynamics and are 
thus integrated out from the action. 
We would like to argue below that one needs to use a  bigger
representation of supersymmetry in order to accomodate
also the glueball degrees of freedom in the effective Lagrangian.  

In order to determine  how glueballs can be included in the action 
let us concentrate our attention on the expression for the $F$ field.
Using the equation  of motion for the gluino  field and for the $D$
component one gets\footnote[5]{In general, one is not allowed to use 
the equation of motion if the VEV of the $F$ field is considered.} 
\begin{eqnarray}
F\equiv \Sigma +i Q \equiv  { \beta (g) \over4  g}[G^2_{\mu\nu}
+i G_{\mu\nu}{\tilde G_{\mu\nu}}]. 
\end{eqnarray}
As we have already mentioned,  
the $F$ field appears in the VY action without
a kinetic term. The term  bilinear in  the $F$ field 
is proportional to 
$
F^+F=\Sigma^2+Q^2. 
$
Besides that,  there are terms linear in  the $F$ field 
in the expression for the effective action, thus, 
the $F$ field can easily be integrated out 
by means of its  algebraic equations of motion \cite {VY}. 

In order to reveal 
subtleties of this procedure let us write down the 
following relation
\begin{eqnarray}
Q={1\over 4! } \varepsilon_{\mu\nu\alpha\beta} H^{\mu\nu\alpha\beta}, 
\label{QC}
\end{eqnarray}
where $ H^{\mu\nu\alpha\beta}$ is a field strength 
for a three-form potential $C_{\nu\alpha\beta}$, 
$
H_{\mu\nu\alpha\beta}=\partial_\mu C_{\nu\alpha\beta}-
\partial_\nu C_{\mu\alpha\beta}-\partial_\alpha C_{\nu\mu\beta}-
\partial_\beta C_{\nu\alpha\mu}.  \nonumber 
$
The $C_{\mu\nu\alpha}$ field  itself is defined as a color singlet   
composite operator of colored gluon fields $A^a_\mu$, 
$
C_{\mu\nu\alpha}={\beta (g) \over 64 g \pi^2}(A^a_\mu 
{\overline {\partial}}_\nu
A^a_\alpha-A^a_\nu {\overline {\partial}}_\mu A^a_\alpha-A^a_\alpha 
{\overline {\partial}}_\nu
A^a_\mu+ 2 f_{abc}A^a_\mu A^b_\nu  A^c_\alpha), \nonumber
$
with $f_{abc}$ being  structure constants of the corresponding $SU(N_c)$
gauge group.  The right-left  derivative in this expression acts  as 
$A{\overline {\partial}}B\equiv A (\partial B)-(\partial A) B $
\footnote[8]{The quantity $Q$ 
can also be expressed through the Chern-Simons
current $K_\mu $ as   $Q=\partial_\mu K_\mu $. Using 
this equation  one can deduce  the relation between the Chern-Simons
current and the three-form potential $C_{\nu\alpha\beta}$, these two
quantities are  Hodge dual to each other:
$K^{\mu}={1\over 3!}\varepsilon^{\mu\nu\alpha\beta}C_{\nu\alpha\beta}$.}.
 
Using these definitions 
one finds that the expression bilinear in  the $F$ field  acquires  the 
following form 
\begin{eqnarray}
F^+F=\Sigma^2-{1\over 4! }  H^2_{\mu\nu\alpha\beta}. \nonumber 
\end{eqnarray}
The second term in this expression is a 
kinetic term  for the three-form potential $C_{\mu\nu\alpha}$.
As before,  the $\Sigma$ field can be integrated out, 
however one should   be
careful in dealing with the $C_{\mu\nu\alpha}$ field. 

In ref. \cite {Gabad} it was  argued that  the 
three-form field $C_{\mu\nu\alpha}$ plays an important role 
in the description of the  pseudoscalar glueball. 
The glueball can be coupled to the QCD $\eta'$ meson by means of the 
$C_{\mu\nu\alpha}$ field \cite {V}.  In the case of SYM theory 
the analog of the $\eta'$ meson is the gluino-gluino 
bound state which acquires mass due to the anomaly in the 
$U(1)_R$ current within the VY approach. 
Thus,  it is natural to  attempt to 
couple the pseudoscalar glueball to the pseudoscalar gluino-gluino 
bound state within the VY action 
using the three-form potential $C_{\mu\nu\alpha}$. 

To elaborate this approach,   let us rewrite the SUSY transformations 
for the components of the $S$ superfield in terms of $\Sigma $
and $C_{\mu\nu\alpha}$ (instead of $F$ and $F^+$) \cite {Gates}
\begin{eqnarray}
\delta_{\zeta} A= \sqrt{2} \zeta \Psi,~~~~~~~~~
\delta_{\zeta}\Psi =i\sqrt{2}\sigma^{\mu}{\bar \zeta}\partial_\mu A
+\sqrt{2}\zeta (\Sigma +{i\over 6}\varepsilon_{\mu\nu\alpha\beta}
\partial^\mu C^{\nu\alpha\beta}), \nonumber \\
\delta_{\zeta}\Sigma = {i\over \sqrt{2}}({\bar \zeta}{\bar \sigma}^\mu
\partial_\mu\Psi+ \zeta \sigma ^\mu \partial_\mu {\bar \Psi}),
~~~~~
\delta_{\zeta}C_{\nu\alpha\beta}={1 \over \sqrt{2}}
\varepsilon_{\nu\alpha\beta\mu }({\bar \zeta}{\bar \sigma}^\mu
\Psi- \zeta \sigma ^\mu  {\bar \Psi}). \nonumber
\end{eqnarray} 
The set of fields given  above forms  an irreducible
representation of supersymmetry algebra. All these
fields can be assigned to a supermultiplet 
introduced in ref. \cite {Gates}. 
That supermultiplet is called a constrained three-form supermultiplet 
\cite {Gates}. 
The easiest way to present this multiplet
is to introduce the following real tensor 
superfield $U$ 
\begin{eqnarray}
U=B+i\theta z  -i {\bar \theta} {\bar z }+{1\over 16}\theta^2 {A^*}+
{1\over 16} {\bar \theta}^2 A+{1\over 48 }\theta \sigma^\mu {\bar
\theta} \varepsilon_{\mu\nu\alpha\beta}C^{\nu\alpha\beta}+ 
\nonumber \\
{1\over 2} \theta^2 {\bar \theta} \left ( {\sqrt{2} \over 8}{\bar
\Psi} +{\bar \sigma}^\mu \partial_\mu z  \right )+
{1\over 2}{\bar  \theta}^2 \theta  \left ( {\sqrt{2} \over 8}
\Psi - \sigma ^\mu \partial_\mu {\bar z  }\right )+{1\over 4}
\theta^2 {\bar \theta^2} \left ( {1\over 4} \Sigma -\partial^2 B\right ).
\label{U}
\end{eqnarray} 
It is a matter of a straightforward calculation to check that the 
real superfield $U$  satisfies the relation\footnote{Despite a 
seeming similarity, the tensor multiplet $U$ should not be
interpreted as a usual vector multiplet. The vector field which might
be introduced in this approach as a Hodge dual of  the three-form 
potential $C_{\mu\nu\alpha}$ would give
mass terms with the wrong sign in our approach (see section 2),
thus,  the actual physical variable  is the three-form potential  
$C_{\mu\nu\alpha}$ rather than its dual vector field 
(the Chern-Simons current).} 
\begin{eqnarray}
{\bar D}^2 U=-{1\over 4} S. 
\label{US}
\end{eqnarray} 
Thus, the real tensor multiplet $U$, defined by the 
expression (\ref {US}),  includes  all the 
components of the chiral supermultiplet $S$. 
Consequently, using the relation (\ref {US}) 
the VY action can be rewritten
in terms of the bigger multiplet $U$.
In addition, the multiplet has also a 
scalar $B$ and  fermion $z $. 
We will show below that this 
allows one to include glueball operators in the effective action. 

First, let us  notice some  features of 
SUSY transformations of the components of the $U$ field. The
components which are shared by the tensor multiplet $U$  
and the chiral
multiplet $S$ (namely $A,~\Psi,~\Sigma~~ {\rm and}~~C$ ) transform 
among themselves, while  other fields ($B$
and $z $) are connected by SUSY rotations to the other four components. 
Furthermore, one can define  a  ``gauge'' transformation of 
the $U$ field as the following shift $U\rightarrow U+Y$, where the superfield
$Y$ satisfies the relation ${\bar D}^2 Y=0$. 
It is important to notice that by means of 
this `` gauge '' transformation one can get rid of the 
$B$ and $z $ fields in the expression for the $U$ multiplet.
This is the analog of the Wess-Zumino gauge for the tensor multiplet $U$. 
Thus, any Lagrangian
written  in terms of the $S$ field only, if reexpressed
 in terms of the $U$ field,
is necessarily invariant under the ``gauge'' 
transformation defined above. 
As a result, the $B$ and $z $ fields can  always be ``gauged''
away from that Lagrangian. Thus in order to be able to retain 
the $B$ and $z $   fields  as dynamical variables  
one must include terms in the Lagrangian which 
breaks this ``gauge'' invariance. The simplest term of this type is the
quadratic term $(U^2)|_D$. Once such a term is included in the
Lagrangian, the ``gauge'' symmetry becomes explicitly broken and the 
$B$ and $z $  components of the superfield $U$ survive as 
dynamical variables. 

Let us now apply the $U$ field formalism to the VY action.
In the case at hand  the chiral symmetry is spontaneously broken by
the gluino condensate. In terms of the chiral superfield this
corresponds to the existence of a nonzero VEV of the $S$ field
$ \langle S\rangle =\mu^3.$
With that in mind the  appropriate relation between the $U$ field and 
the chiral multiplet is
\begin{eqnarray}
{\bar D}^2 U=-{1\over 4} (S-\langle S\rangle ). 
\label{U_S}
\end{eqnarray}
The only result of this modification is that the field $A$
in eq. (\ref {U}) gets replaced by the quantity 
$A-\langle A \rangle$. 

Now use the relation (\ref {U_S})
to write the action  
in terms of the $U$ field.
In order to break the ``gauge'' invariance of the VY action we
add  a  term proportional to $U^2$ to the VY Lagrangian. 
An  appropriate  term with zero  R-charge and correct  
dimensionality  is
\begin{eqnarray}
\left ( -{U^2\over (S^{+}S)^{1/3}} \right )|_D.
\end{eqnarray}
Below, we are going to show that 
once this term is added to the VY action,
the following fields become dynamical:
\begin{itemize}

\item The $B$ field propagates and it represents
one massive real scalar degree of freedom
(identified later with the scalar glueball).

\item The three-form potential $C_{\mu\nu\alpha}$, which becomes massive,
also propagates. It represents one physical  degree of freedom 
(identified with the pseudoscalar glueball). 

\item The complex field $A$,  being 
decomposed into parity eigenstates,   describes the massive gluino-gluino 
scalar,  $s$, and pseudoscalar, $p$, mesons.

\item $z $ and $\Psi$ describe
the massive gluino-gluon fermionic bound states. 

\end{itemize}
Relations between
masses of these states will be given in the next section.  

Based on the arguments given abowe one can write
down the effective Lagrangian  for the lowest-spin multiplets 
of the $N=1$ SUSY YM theory in the following form
\begin{eqnarray}
{\cal L}={1\over \alpha} (S^{+}S)^{1/3}|_D+ 
\gamma [( S \log {S\over \mu^3}-S)|_F+{\rm h.c.}]+   
{1\over \delta} \left ( -{U^2\over (S^+S)^{1/3}} \right )|_D,
\label{NewA}
\end{eqnarray}
where $\alpha~{\rm and}~\delta$ are arbitrary positive constants. 
One obtains the VY Lagrangian in the limit $\delta \rightarrow \infty$.
In general, higher powers of $U$  
can also be added to this Lagrangian.
Those terms would introduce new quartic, quintic  and other  higher 
interaction terms. However, the quadratic part of the action which
defines two-point Green's functions and masses will not be affected. 
In that respect, the effective Lagrangian (\ref {NewA}) 
can be considered  as the  one describing small 
perturbations of fields about a vacuum state. 

Let us determine the SUSY vacuum state defined by the 
Lagrangian (\ref {NewA}). The potential for the model is 
a complicated function of the variables present in the $U$ 
superfield. After integration over the auxiliary $\Sigma$ field,
the bosonic part of the potential is  
\begin{eqnarray}
V_0= {2\over \delta (16)^2} {|\phi |^6+\mu^6-2\mu^3|\phi |^3 {\rm cos} 3\rho
\over |\phi |^2}+ {3 \over \delta (48)^2}{ C^2_{\mu\nu\tau}
\over |\phi |^2}+ \nonumber \\
+{9\alpha |\phi |^4\over 4} {1\over 1- {\alpha\over \delta} 
{B^2\over |\phi |^4}} \left ({B\over 24  \delta |\phi |^2}
\{1+ { 2 \mu^3\over |\phi |^3}{\rm cos} 3\rho \}-3\gamma \log {|\phi |^2\over
\mu^2}\right )^2,    
\label{Potential}
\end{eqnarray}
where we introduced the notations  $\phi\equiv A^{1/3}$ and 
$\rho\equiv {\rm arg}\phi$. 

In order to find the vacuum state one
should find the absolute minimum of the potential (\ref {Potential}).
Since we are dealing with a supersymmetric model, the value
of the potential in that minimum has to be zero. 
As a  result of Lorentz invariance,  the VEV of the three-form field 
is zero, i.e. $\langle C_{\mu\nu\tau} \rangle=0$. 
The VEV of $Q$ is also zero due to the CP invariance of
the model. 
After some algebra one finds that
the only global,  CP invariant  minimum of the potential (\ref {Potential})
is given by:
$\langle \phi \rangle =\mu$, 
$\langle B \rangle =\langle C \rangle = \langle \rho \rangle=0$. 
The effective Lagrangian (\ref {NewA}) describes small 
perturbations of fields about the vacuum state defined by these VEV's.  

We would like to make a comment  here. 
The singularity in the potential at $\delta |\phi|^4 =\alpha B$
indicates that for large perturbations the higher dimensional
terms omitted in  (\ref {NewA}) should become important.
The same 
multiplier $1- {\alpha\over \delta} 
{B^2\over |\phi |^4}$ appears in front of kinetic terms 
for the scalar fields, so the physical potential is always 
positively defined. 
As we mentioned above,  we are mainly concerned with the mass 
spectrum of the model which can be studied using small perturbations 
about the ground state, so that the approximation 
given in (\ref {NewA}) is good enough  for our goals. 

The actual physical states described by the action are 
mixed states. Below, we deduce the masses of these mixed physical states.
Let us write down  the mass and mixing terms of the
Lagrangian separately. 
One finds the following pairs of bosonic variables being mixed with one
another
\begin{eqnarray} 
B - s ~~ {\rm system}:~~~ {9 \alpha \over \delta (16)^2}\mu^2 s^2+
{81 \over 2} \alpha^2 \gamma^2 \mu^2 s^2+{9 \alpha \over \delta 
(16)^2}\mu^2 B^2-
{27 \sqrt {2}\over 16} \sqrt {\alpha \over \delta}\alpha \gamma \mu^2 Bs;
\nonumber \\
C - p ~~ {\rm system}:~~~~{9 \alpha \over \delta (16)^2 }\mu^2 p^2+
{27 \alpha\over (48)^2 \delta}\mu^2 C_{\mu\nu\tau}^2+
{9\sqrt{2}\over 6}\alpha \gamma  \mu p \varepsilon_{\mu\nu\tau\sigma}
\partial^\mu C^{\nu\tau\sigma};~~~~~~~~~~
\end{eqnarray}

In order to find the  physical masses one must 
diagonalize the corresponding mass matrices. 
Concentrate, for instance, on the first row of these expressions 
which describes the mixed state of scalar meson, $s$, and scalar
meson-glueball, $B$. The former gets mass both from the superpotential and
$U^2$-term while the latter gets mass only from the $U^2$-term.  
When the mixing
term is switched on, the initially heavier state ($s$) gets even  heavier,
and the initially lighter state ($B$) becomes even lighter than they were
originally.  Performing the diagonalization, 
one finds that  the physical eigenstates are  mixed states with 
the following mass eigenvalues
\begin{eqnarray}
{1\over 2} M^2_{\pm }={9 \alpha \over \delta (16)^2 }\mu^2+
{81 \over 4} \alpha^2 \gamma^2 \mu^2 \pm {81 \over 4} \alpha^2 \gamma^2 \mu^2
\sqrt{ 1+ {1 \over 288}{ \alpha\over  \delta} {1\over
(\alpha\gamma)^2} }. 
\label{up}
\end{eqnarray}
Here, the subscript ``$+$'' refers to the heavier state ${\tilde s}$
which, without
mixing,  would have been a 
 pure gluino-gluino bound state (the $s$ particle). The subscript 
``$-$'' refers to the lighter state ${\tilde B}$  ($B$ in the absence 
of mixing). 

Studying the potential of the model, we find  that
the physical eigenstates fall into the  two different 
``multiplets".  Neither of them contain pure gluino-gluino, gluino-gluon or
gluon-gluon bound states.  Instead, the physical excitations are 
mixed states of these composites.  The heavier set of states contains: 
\begin{itemize}
\item A pseudoscalar meson, which without mixing reduces to the $0^{-+}$
($l=0$)
gluino-gluino bound state (the analog of the QCD $\eta'$ meson).
\item A scalar meson that without mixing is an $0^{++}$ ($l=1$)
gluino-gluino excitation.
\item A mixed fermionic gluino-gluon bound state. 
\end{itemize}
These heavier states become the chiral supermultiplet described by the VY
action in the limit that the additional term we have added to the effective
Lagrangian is  removed.  The new states which appear as a result of our 
generalization forms a lighter multiplet:
\begin{itemize}
\item A scalar meson, which for small mixing becomes a $0^{++}$ ($l=0$)
glueball. 
\item A pseudoscalar state, which for small mixing is identified as a
$0^{-+}$ ($l=1$) glueball.
\item A mixed fermionic gluino-gluon bound state. 
\end{itemize}

We call the reader's attention to an interesting feature of the
effective action introduced here.  Although the physical states fall into
multiplets whose $J^P$ quantum numbers 
correspond to two chiral supermultiplets,
the action is written in terms of one real tensor supermultiplet $U$. 
The natural question arises whether the whole action can be 
rewritten in terns of two different chiral multiplets. 
The relation between a real tensor and chiral
supermultiplets  (the so called chiral-linear duality) 
was established in ref. \cite {Gates}.
For SYM theory the chiral-linear duality was used in refs. \cite 
{Derendinger}. 
Applied to our problem the results  of refs. \cite
{Gates} and \cite {Derendinger}
can be stated as follows. One introduces into the effective 
Lagrangian a new chiral superfield, let us denote it by  $\chi$
\begin{eqnarray}
\chi (y, \theta) \equiv \phi_{\chi}(y)+ \sqrt {2}\theta
\Psi_{\chi}(y)+\theta^2 F_{\chi}(y).
\label{chi} 
\end{eqnarray}
One can find an  effective Lagrangian written in terms
of two chiral superfields,  $S$ and $\chi$  which is 
equivalent to the expression given in (\ref {NewA}). 
In our case 
\begin{eqnarray}
{\cal L}~=~{1\over \alpha}\, (S^{+}\,S)^{1/3} \Big|_D ~+~ 
{\delta \over 4 }\, (S^{+}\,S)^{1/3}\, (\chi~+~ \chi^{+})^2 \Big|_D
~+~ \nonumber \\
\Big[\gamma \,( S \,\log {S\over \mu^3}~-~S)\Big|_F ~+~ 
{1 \over 16} \,\chi\, (S ~-~ \mu^3)\Big|_F~+~{\rm h.c.}\Big]~.
\label{2s}
\end{eqnarray}
Comparing this expression to the VY Lagrangian one 
notices that both the K{\"a}hler potential and the superpotential
are modified by new terms. The  multiplets $S$ and $\chi$ are independent.

We would like to relate  this expression to 
the Lagrangian of the theory written in terms of the $U$ field (\ref {NewA}). 
If the $U$ field is postulated  as a fundamental degree of freedom,
then the $S$ field is a  derivative superfield $ S= \mu^3-4 {\bar D}^2 U$.
Using this relation 
the Lagrangian (\ref {2s}) can be rewritten as 
\begin{eqnarray}
{\cal L}~=~{1\over \alpha}\, (S^{+}\,S)^{1/3} \Big|_D ~+~ 
{\delta \over 4 }\, (S^{+}\,S)^{1/3}\, (\chi~+~ \chi^{+})^2 \Big|_D
~+~ \nonumber \\
\Big[\gamma \,( S \,\log {S\over \mu^3}~-~S)\Big|_F ~+~{\rm
h.c.}\Big]~+~ U (\chi~+~ \chi^{+})\Big|_D~.
\label{2cs}
\end{eqnarray}
This expression depends on two superfields $U$ and $\chi$ ($S$ 
is expressed through $U$ in accordance with (\ref {U_S})). However, 
the dependence on the chiral superfield $\chi$ is trivial, 
the combination  $\chi +\chi^{+}$ can be integrated out 
from the Lagrangian  (\ref {2cs}). As a result one derives 
\begin{eqnarray}
\chi + \chi^{+}= -{ 2U\over \delta (S^{+}S)^{1/3}}.
\label{Uchi}
\end{eqnarray}
Substituting this expression back into the Lagrangian
(\ref {2cs}) one arrives at  the original  expression (\ref {NewA})
where the $S$ field is a derivative field satisfying 
the relation (\ref {U_S}).

Let us stress again that the descriptions in terms of the Lagrangian 
(\ref {NewA}) and (\ref {2s}) are equivalent on the mass-shell. 
In the Lagrangian (\ref {NewA}) the dynamical degrees of freedom are 
assigned to  the only superfield $U$, while in the Lagrangian  (\ref
{2s}) the physical degrees of freedom are found  as components of two
chiral  supermultiplets $S$ and $\chi$. The peculiarity of 
the expression (\ref {2cs}) is that the chiral superfield 
$\chi$ enters only through the real combination  $\chi+\chi^+$.
That is why it was possible  to formulate the action in terms only 
of the real superfield $U$.  It is essential from a physical point of
view since the component glueball field must be real.

In order to make contact with the results of lattice simulations of
SUSY YM model \cite {Lattice} one needs to consider the 
model with soft SUSY breaking term introduced via gluino mass.  

The potential of the softly broken model 
consists of two parts , $V_0$ defined in (\ref
{Potential}) and an additional  SUSY breaking term \cite {FGS2}
\begin{equation}
V=V_0+ \widetilde{m}_\lambda \,Re \Big({\mu^3 \over 16}\, \phi_\chi
~+~ \gamma \,\phi^3 \Big),  \nonumber
\end{equation}
where 
$\widetilde{m}_\lambda \equiv {32 \pi^2 \over g^2  N_c}\, m_\lambda$.

One calculates minima of  the full scalar potential $V$.
Explicit  though tedious  calculations yield the following results.
The VEV of the $\phi$ field does not get shifted
when the soft SUSY breaking terms are introduced.  Thus, even  in the 
broken theory $\langle \phi   \rangle =\mu $. However,  
the $\phi_\chi$  (and $B$ ) fields  acquire  nonzero VEVs 
in the broken case
\begin{eqnarray}
\langle \phi_\chi \rangle  = 
{8 \over 9 \alpha \mu} \widetilde{m}_\lambda ~~~~~{\rm and }~~~~~~
\langle B \rangle=-{8 \delta \over 9 \alpha }\widetilde{m}_\lambda \mu .
\label{VEV}
\end{eqnarray}
The shift of the vacuum energy  causes  the spectrum of the model to be 
also rearranged. Explicit calculations of the masses of all 
lowest-spin states yield  the following results
\begin{equation}
\label{mscalar}
M^2_{scalar\,\pm}~ ~=~~ M^2_{\pm} ~-~ 
{3 \over 4}\, \alpha \gamma\, \mu \,\widetilde{m}_\lambda \Bigg (
1~ \pm ~ \sqrt {1+x}~ \Bigg ) \Bigg (2~\pm~ {1\over \sqrt {1+x}}~
\Bigg),
\end{equation}

\begin{equation}
\label{mfermion}
M^2_{fermion \,\pm} ~=~M^2_{\pm} ~-~
{3 \over 4}\, \alpha \gamma\, \mu \,\widetilde{m}_\lambda \Bigg(
1 \pm \sqrt {1+x}~ \Bigg ) \Bigg (3 ~\pm~ {1\over \sqrt {1+x}}~
\Bigg),
\end{equation}

\begin{equation}
\label{mpseudo}
M^2_{p-scalar \,\pm} ~=~M^2_{\pm} ~-~
{3 \over 4}\, \alpha \gamma\, \mu \,\widetilde{m}_\lambda \Bigg (
1 \pm \sqrt {1+x} ~ \Bigg ) \Bigg (4~\pm~ {1\over \sqrt {1+x}}~
\Bigg),
\end{equation}
where $M^2_{\pm}$ denote the masses in the theory with unbroken SUSY
\cite {FGS2} and  $ x\equiv {1 \over 288}
{\alpha \over \delta} {1\over(\alpha \gamma)^2}.$
In these expressions  the plus sign refers to the heavier
supermultiplet and the minus sign to the lighter set of states. 
One can verify that these values satisfy the mass sum rule to  
leading order in $O(m_{\lambda})$: 
\begin{eqnarray}
\label{sum}
\sum_{j} \,(-1)^{2j+1}\, (2j+1)\, M_j^2~=~ 0~,
\nonumber
\end{eqnarray}
where the summation goes over the spin $j$ of 
particles in the supermultiplet.

Let us discuss the mass shifts given in eqs. 
(\ref{mscalar}-\ref{mpseudo}). Consider 
the light  supermultiplet. In accordance with eqs. 
(\ref{mscalar}-\ref{mpseudo}), the masses in the light  multiplet are
increased  in the broken theory.  The biggest  mass shift is 
found in the pseudoscalar channel. The smallest shift  is observed 
in the scalar channel. The fermion mass falls in  between these two
meson states. Thus, the lightest state in the 
spectrum of the model is the particle 
which without mixing would have been the scalar glueball.
There is a fermion state above that scalar. Finally, the
pseudoscalar glueball is heavier than those two states. 

Let us now turn to the heavy supermultiplet. In the broken 
theory the masses in  that multiplet get pulled down. However,
all states of the heavy multiplet are still heavier than any state
of the light multiplet in the domain of validity of our approximations.
The ordering of the states in the heavy supermultiplet is just the
opposite as in the light supermultiplet: the lightest state is the
pseudoscalar meson, the heaviest is the scalar, and the fermion, as
required, falls between them.  
The qualitative features of the  spectrum are shown in fig. 1. 
\vspace{0cm}
\epsfysize=24.0 cm
\begin{figure}[htb]
\epsfbox{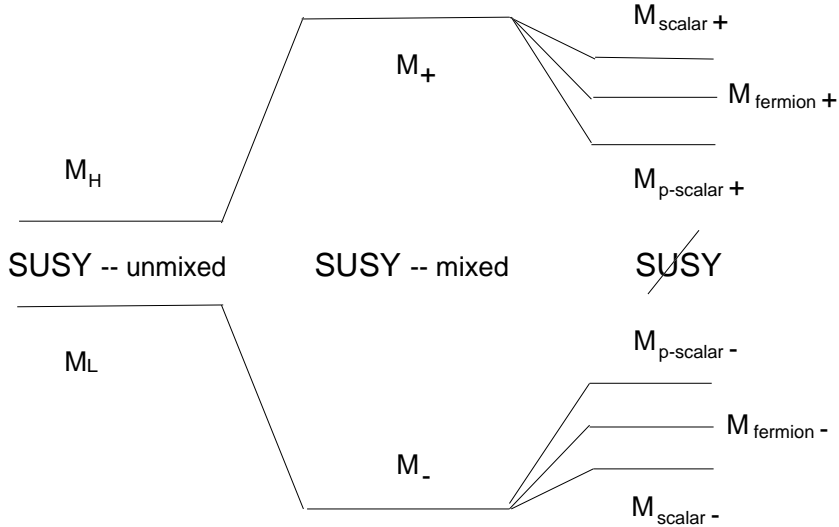}
\vspace{-16.5cm}
\caption{ Qualitative behavior of mass spectrum when passing from SYM to
softly broken model.
}
\label{phasediag}
\end{figure}

It is not surprising that the lowest mass state  obtained in
(\ref{mscalar}-\ref{mpseudo}) is a scalar particle. 
This is in  agreement with the result  of
ref. \cite{West} where  it was shown that the mass of the lightest
state which couples  to the operator $G_{\mu\nu}^2$ is less than the
mass of the  lightest state that couples  to $G \tilde{G}$, in pure
Yang-Mills theory. As a result, the lightest glueball turns out to be
the scalar glueball \cite{West}.  One can  apply the method  of ref.
\cite{West} to the SYM theory as well.  Due to the positivity of the
gluino determinant (see ref. \cite{steve}) one also deduces  that the
lightest state in softly broken SYM spectrum should be a scalar
particle. The pseudoscalar of that multiplet is therefore heavier.

Our result that the multiplet containing glueballs is split in such a
way that the scalar is lighter than the pseudoscalar, and vice versa
for the multiplet containing gluino-gluino bound states, is consistent
with expectations from quark-model lore.  In ordinary mesons the
$l=1$ states are heavier than their $l=0$ counterparts and the $l=0$
gluino-gluino bound state is a pseudoscalar, while an $l=0$
gluon-gluon bound state is a scalar.  It is interesting that in SYM
with massless gluinos the $l=0$ and $l=1$ bound states are degenerate,
but when the gluino masses are turned on one recovers the expected
ordering seen in $q \bar q $ states.

Summarizing, we have shown that the generalized VY effective action
can be written in two different ways. In one case the fundamental
superfield upon which the action is constructed is  the real 
tensor superfield $U$. In another approach all degrees
of freedom of the model are described by two chiral superfields
$\chi$ and $S$. In both cases the spectrum consists 
of two multiplets which are not degenerate in masses
even when SUSY is unbroken. 
The spin-parity quantum numbers  of these multiplets are identical to 
those of certain  chiral supermultiplets. 

We introduced a  soft SUSY breaking term in the 
Lagrangian of the $N=1$ SUSY Yang-Mills model. The spurion method was used
to calculate the corresponding soft SUSY breaking terms in the generalized
VY Lagrangian. These soft breaking terms cause a shift of the vacuum energy 
of the model. The physical eigenstates which are  degenerate in the SUSY
limit, are  split when  SUSY breaking is introduced.  
We studied these mass splittings in detail. 
We have confirmed  that the spectrum of the broken theory
is in agreement with some low-energy theorems \cite{West}, namely  
the scalar glueball turns out to be lighter than the pseudoscalar one.
The results of the present paper   can be directly tested in lattice
studies  of $N=1$ supersymmetric Yang-Mills theory \cite {Lattice}.

Even when SUSY is unbroken,  the physical mass eigenstates are not pure 
gluon-gluon, gluon-gluino or gluino-gluino
composites; rather, the physical particles are mixtures of them.
The multiplet which without mixing would have been  
the glueball multiplet is lighter. As a result, those states cannot be 
decoupled  from the effective Lagrangian. This means that
comparisons  of lattice results to  analytic predictions
based on the original VY action are not justified.  

The work was partially supported by grant No. NSF-PHY-94-23002.

\end{document}